\documentclass[11pt]{article}
\usepackage{latexsym}

\begin{document}
\title{Generalized protocol for distribution and concentration of Quantum information }
\author{Partha Mukhopadhyay \\ Cryptololgy Research Group \\Applied Statistics Unit \\ Indian Statistical Institute \\ Kolkata-700108 \\ India}
\date{22.08.2003}
\maketitle

\begin{abstract}
Alice can distribute a quantum state $|\phi\rangle$ to $N$
spatially separated parties(say Bobs) by telecloning. It is
possible for Charlie to reconstruct the quantum state to him if he
shares same kind of telecloning quantum channel with Bobs using
only LOCC. For $N=3$ reconstruction can be done faithfully using
Smolin's 4 party unlockable bound entangled state as shared
channel. In this note we investigate, in multiparty setting, the
general structure of quantum channel and protocol by which
faithful distribution and concentration of quantum information can
be done.
\end{abstract}


\maketitle

\section{Introduction}

\vspace{0.2cm}

In the usual quantum teleportation scheme \cite{bennett93} an
unknown quantum state can be faithfully transmitted to a remote
receiver via an initially shared maximally entangled state between
sender and receiver. This can be considered as basic unit of
quantum communication network. But in the case of distributed
multiparty communication information is transferred usually in
the following way. Firstly the unknown state is distributed among
many distant spatially separated intermediate receivers via an
initially shared channel between them. Then the original state is
remotely concentrated back to the actual receiver by applying
Local Operations and Classical Communication (LOCC) by the
intermediate receivers via another channel. In both distribution
and concentration we require initially shared entangled state as
channel state among the parties in the network. In a recent work
\cite{murao99} it was shown that an unknown state of a $2$-level
system (called qubit) can be distributed among $N$ distant qubits
such that each qubit is either optimally cloned or anti-cloned.
This process is called telecloning. As an example a sender (Alice,
say) first distributes an unknown qubit $|\phi \rangle = \alpha|0
\rangle + \beta|1 \rangle$ among three spatially separated distant
qubits ($M_a$, $M_o$, $M_c$) and after distribution the state
becomes

\begin{eqnarray}
|\psi_D \rangle_{M_a M_o M_c}  = & \alpha \sqrt{\frac{2}{3}} \left
\{ |0\rangle |00
\rangle + \frac{1}{2} |1\rangle (|01\rangle + |10\rangle) \right \} + \nonumber \\
& \beta \sqrt{\frac{2}{3}}\left \{ |1\rangle |11\rangle +
\frac{1}{2} |0\rangle (|01\rangle + |10\rangle) \right \}
\end{eqnarray}

where the first qubit is the anticloned and the last two are
cloned ones. Now this distributed state can again be concentrated
to another single qubit (holding by Bob, say) using Smolin's four
qubit unlockable bound entangled state given by
$\frac{1}{4}(P[\phi^+] + P[\phi^-] + P[\psi^+] + P[\psi^-])$
shared between parties holding ($M_a$, $M_o$, $M_c$)and Bob
(Smolin state \cite{smolin})
 using only  LOCC \cite{murao01}. $\phi^+ , \phi^- , \psi^+ , \psi^-$ are Bell's state.

 \vspace{0.3cm}

 In this note we investigate a general class of $N$ party pure and mixed
 state which can be used for faithful distribution and
 concentration of qubit using only LOCC. We also propose the
 protocols used for distribution and concentration. We observe
 that telecloning state and Smolin's state belong to the
 generalized class.

\vspace{0.5cm}

\section{Distribution and concentration of qubit}

In this section, we will show how an unknown state can be
distributed among several intermediate spatially separated parties
and then again remotely reconstructed to another distant party
without any global operation.

Suppose we have (N+1) parties where N is odd, and Alice, one among of
them, has a unknown qubit $|\psi\rangle = \alpha|0\rangle +
\beta|1\rangle$ where, $|\alpha|^2 + |\beta|^2 = 1$. Alice 
firstly distributes her qubit among the $N$ Bobs, and Bobs then
reconstruct the original state to another remote party Charlie,
without any global operation.

So the protocol has two phase. We call the first phase as
distribution phase and the second phase as concentration phase.

In the first phase, Alice shares the following state with $N$ Bobs
\begin{eqnarray}
|{\chi}^{(1)}\rangle_{A, B_1,....B_{N}} = &
\left(1/\sqrt{2}\right)[ |0\rangle_A \otimes
\sum_{i}a_{i}|S_{i}\rangle_{B_1,....B_N} \nonumber \\
& + |1\rangle_A \otimes
\sum_{i}a_{i}|\bar{S}_{i}\rangle_{B_1,....B_{N}}].
\end{eqnarray}

Where $|\bar{S}_i\rangle_{B_1,....B_{N}}$ is the qubit wise
complement state of $|S_i\rangle_{B_1,....B_{N}}$. The state is
properly normalized, i.e $\sum_{i}a_{i}\bar{a}_{i}=1$. Where
$\bar{a}$ is the complex conjugate of $a$. In each
$|S_{i}\rangle_{B_1,....B_{N}}$ the no of $|0\rangle_{B_i}$'s is
odd .  Now Alice performs Bell state measurement on her two
qubits, the unknown qubit and one qubit of the shared channel,
where Bell states are
$$|Bell_1 \rangle = |\phi^+\rangle = [|00\rangle +
|11\rangle]/\sqrt{2},$$ $$|Bell_2 \rangle = |\psi^+\rangle =
[|01\rangle + |10\rangle]/\sqrt{2},$$ $$|Bell_3 \rangle =
|\psi^-\rangle = [|01\rangle - 10\rangle]/\sqrt{2}$$ and
$$|Bell_4 \rangle = |\phi^-\rangle = [|00\rangle -|11\rangle]/\sqrt{2}.$$
Depending on her outcome, which is one of the following four
states $\phi^+, \phi^-,\psi^+, \psi^-$, Alice makes phone call to
$N$ Bobs to perform usual unitary operation $I, \sigma_{z},
\sigma_{x}, \sigma_{y}$ respectively on their own qubits i.e each
Bob will individually apply $I$ for measurement result $\phi^+$,
$\sigma_{z}$ for $\phi^-$, $\sigma_{x}$ for $\psi^+$ and
$\sigma_{y}$ for $\psi^-$. After this the state of the $N$ Bobs'
become $|{\xi}^{(1)}\rangle = \frac{1}{k}(\alpha
{\sum}_{i}a_{i}|S_{i}\rangle + \beta
{\sum}_{i}a_{i}|\bar{S}_{i}\rangle)$, where $k$ is the normalizing
factor. We note that the telecloning channel is a member of this
general class channel shared between Alice and Bobs.

Now instead of using $|{\chi}^{(1)}\rangle$, as a channel state,
if we take the channel state as a mixed one, which is the mixture of
any no. of states of the form $|{\chi}^{(i)}\rangle$, i.e. $\rho =
{\sum}_j p_j P[|{\chi}^{(j)}\rangle]$, with ${\sum}_j p_j= 1$
where

\begin{eqnarray}
|{\chi}^{(1)}\rangle_{A, B_1,....B_{N}} = &
\left(1/\sqrt{2}\right)[ |0\rangle_A \otimes
\sum_{i}a_{i}|S_{i}\rangle_{B_1,....B_N} \nonumber \\
& + |1\rangle_A \otimes
\sum_{i}a_{i}|\bar{S}_{i}\rangle_{B_1,....B_{N}}],
\end{eqnarray}

\begin{eqnarray}
|{\chi}^{(2)}\rangle_{A, B_1,....B_{N}} = &
\left(1/\sqrt{2}\right)[ |0\rangle_A \otimes
\sum_{i}b_{i}|S_{i}\rangle_{B_1,....B_N} \nonumber \\
& + |1\rangle_A \otimes
\sum_{i}b_{i}|\bar{S}_{i}\rangle_{B_1,....B_{N}}],
\end{eqnarray} and so on, then after using the same distribution
protocol the Bobs will share a state $\sigma = {\sum}_j q_j
P[|{\xi}^{(j)}\rangle]$ where ${\sum}_j q_j = 1$.

We note that Smolins 4 party bound entangled state belongs to the
mixed state channel for $N=3$ , $p_i$'s are equal and the coefficients of $|S_i\rangle$ (and so for $|\bar{S}_{i}\rangle$) are same for $|{\chi}^{(1)}\rangle$ , $|{\chi}^{(2)}\rangle$ and $|{\chi}^{(3)}\rangle$.

In the concentration phase, Charlie wants to
reconstruct the qubit to him. Charlie shares a
properly normalized state
\begin{eqnarray}
|{\chi}^{(2)}_{{B}^{'}_{1}....{B}^{'}_{N}, C}\rangle = &
(1/\sqrt{2})[({\sum}_{i}b_{i}|S_{i}\rangle) \otimes |0\rangle \nonumber \\
& + ({\sum}_{i}b_{i}|\bar{S}_{i}\rangle) \otimes |1\rangle]
\end{eqnarray}
with $N$ Bobs among whom the original state was distributed, where
${\sum}_{i} b_{i}\bar{b}_{i} = 1$ and $\bar{b}_i$ is complex
conjugate of $b_i$. Now each of the Bob performs Bell measurement
on his two qubits and let Bob$_i$ gets $|{Bell}^{(i)}_{k}
\rangle$. Bobs inform their measurement results to Charlie by phone call. Charlie then performs unitary operation on his qubit given by $_{i\in
\{1,N\}}{\Pi}({\sigma}^{(i)}_{k})$, where ${\sigma}^{(i)}_{k}$ is
$I$, $\sigma_x$, $\sigma_y$ or $\sigma_z$, depending on whether
Bob$_i$'s outcome is $\phi^+$, $\psi^+$ $\psi^-$ or $\phi^-$. For
example if Bob$_1$ gets $|\phi^{+} \rangle $, Bob$_2$ gets
$|\psi^{-} \rangle$ $\cdots$ Bob$_N$ gets $|\psi^{+} \rangle$ then
Charlie will perform $I.\sigma_y.\cdots.\sigma_x$. After the
operation done by Charlie the state $|\psi\rangle$ is formed at
Charlie's end exactly. This protocol holds good for the mixed state channel also.

\vspace{0.3cm}

Interestingly this protocol does not work if $N$ is even. We are
able to propose a general protocol which works for any $N$.

As before, Alice holding a qubit $|\psi\rangle = \alpha|0\rangle +
\beta|1\rangle$ where, $|\alpha|^2 + |\beta|^2 = 1$, wants to
distributes the qubit among $N$ Bobs.

The entangled channel shared among Alice and $N$ Bobs are given
by
\begin{eqnarray}
|{\chi}^{(1)}\rangle_{A, B_1,....B_{N}} = &
\left(1/\sqrt{2}\right)[ |0\rangle_A \otimes
\sum_{i=0}^{n-1}a_{i}|0^{(n-i)}1^i\rangle_{B_1,....B_N} \nonumber \\
& + |1\rangle_A \otimes
\sum_{i=0}^{n-1}a_{i}|1^{(n-i)}0^i\rangle_{B_1,....B_{N}}].
\end{eqnarray}

The state is properly normalized. Now Alice performs Bell state
measurement on her two particles and makes phone calls to $N$
Bobs informing the measurement outcomes.

Now if Alice's outcome is $\phi^+$, each Bob operates $I$ on
their corresponding qubit. If Alice's outcome is $\phi^-$, first
of the intermediate Bobs performs $\sigma_z$ on his qubit and
other performs $I$ on their corresponding qubit.

Now if Alice's outcome is $\psi^+$, each Bob operates $\sigma_x$
on their corresponding qubit. If Alice's outcome is $\psi^-$,
first of the intermediate Bobs performs $\sigma_y$ on his qubit
and other performs $\sigma_x$ on their corresponding qubit.

So after the distribution the states between $N$ Bobs will be
given by $|{\xi}^{(1)}\rangle = \frac{1}{k}(\alpha
{\sum}_{i=0}^{n-1}a_{i}|0^{(n-i)}1^i\rangle + \beta
{\sum}_{i=0}^{n-1}a_{i}|1^{(n-i)}0^i\rangle)$, where $k$ is the
normalizing factor.

\vspace{0.1cm}

Instead of using pure state channel if we use mixed state channel
then the result will be of same nature as in the last protocol.
\vspace{0.2cm}

Let us now consider the distribution phase. Charlie, who wants to
reconstruct the qubit to him, shares properly normalized entangled
channel
\begin{eqnarray}
|{\chi}^{(2)}\rangle_{B_1,....B_{N},C} = &
\left(1/\sqrt{2}\right)[
\sum_{i=0}^{n-1}b_{i}|0^{(n-i)}1^i\rangle_{B_1,....B_N}\otimes |0\rangle_C\nonumber \\
&
+\sum_{i=0}^{n-1}b_{i}|1^{(n-i)}0^i\rangle_{B_1,....B_{N}}\otimes|1\rangle_C].
\end{eqnarray}
with $N$ Bobs'.

Now each Bob performs bell state measurement on his two
qubit and informs Charlie the outcome of their respective
measurement.

Now Charlie can reconstruct the qubit if he follows the following
Algorithm:

$1$. Charlie initialize a counter $S$ to zero.

$2$. Whenever Charlie finds the measurement outcome of any Bob
$\phi^-$ or $\psi^-$, he increases $S$ by one.

$3$. Charlie stores the measurement outcome of $B_1$.

\vspace{0.2cm}

Now, after knowing measurement outcomes from each Bob
Charlie follows the following set of rules to reconstruct the
qubit:

$1$. If $S$ is even then Charlie can reconstruct the distributed
qubit to him by operating  $I$ or $\sigma_z$ or $\sigma_x$ or
$\sigma_y$ on his qubit whenever the corresponding measurement
outcome of $B_1$ is $\phi^+$ or $\phi^-$ or $\psi^+$ or $\psi^-$
respectively.

$2$. If $S$ is odd then Charlie can reconstruct the distributed
qubit to him by operating  $\sigma_z$ or $I$ or $\sigma_y$ or
$\sigma_x$ on his qubit whenever the corresponding measurement
outcome of $B_1$ is $\phi^+$ or $\phi^-$ or $\psi^+$ or $\psi^-$
respectively.

\vspace{0.2cm} If we consider any mixture of the pure state as
channel (as in the $1^st$ protocol) and follow the same protocol
then also faithful reconstruction of the qubit can be done.

\vspace{0.5cm}

\section{Conclusion}
\vspace{0.3cm} We have proposed a large class of pure and mixed
state multiparty quantum channel which can be faithfully used for
distribution and concentration of quantum information. In the case of
first protocol, we observed that telecloning channel and Smolin's state channel are the special cases of our pure and mixed state channel
respectively. But surprisingly we could not extend the protocol
for $N$ even.

In the second case we are able to present a class of channels and
corresponding protocol which can be used for any $N$. It remains
an interesting problem to find, what will be the most general
channel and protocol that can be used for any $N$.

\vspace{0.4cm} Author acknowledges Guruprasad Kar, Sibasish Ghosh,
Anirban Roy of Indian Statistical Institute(India) and Debasis Sarkar of
Calcutta University (India) for many useful suggestion and helpful
discussion during this work.

\vspace{0.4cm}

\small{ \noindent{E-mail} : partha\_isi@yahoo.com }


\vspace{0.2cm}

\begin{thebibliography}{99}
\bibitem{bennett93} C.H. Bennett, G. Brassard, C. Crepeau, R. Jozsa, A. Peres and
W.K Wotters, \textit{Phys. Rev. Lett.} \textbf{70} 1895 (1993).
\bibitem{murao99} M. Murao, D. Jonathan, M. B. Plenio and V.
Vedral \textit{Phys. Rev. A} \textbf{59} 156 (1999);
quant-ph/9806082.
\bibitem{smolin} J.A. Smolin, \textit{Phys. Rev. A} \textbf{63} 032306
(2001); quant-ph/0001001.
\bibitem{murao01} M. Murao and V. Vedral \textit{Phys. Rev. Lett.}
\textbf{86} 352 (2001); quant-ph/0008078.
\bibitem{brun} Todd A. Brun, quant-ph/0102046.
\end{thebibliography}
\end{document}